\documentclass[submission,copyright,creativecommons]{eptcs}
\providecommand{\event}{IMPEX 2017 and FM\&MDD 2017} 
\usepackage{breakurl}             

\title{A Formal Model to Facilitate Security Testing in\\Modern Automotive Systems}
\author{Eduardo dos Santos \qquad Andrew Simpson
\institute{Cyber Security Centre for Doctoral Training\\
Department of Computer Science\\
University of Oxford\\
Oxford, OX1 3QD, United Kingdom}
\email{eduardo.dossantos@cs.ox.ac.uk \quad andrew.simpson@cs.ox.ac.uk}
\and
Dominik Schoop 
\institute{Esslingen University of Applied Sciences\\
73732 Esslingen am Neckar, Germany}
\email{\quad Dominik.Schoop@hs-esslingen.de}
}
\def\titlerunning{A Formal Model for Automotive Security Testing}
\def\authorrunning{E. dos Santos, A. Simpson \& D. Schoop}

\usepackage{color}
\usepackage{amsmath}
\usepackage{amssymb}

\usepackage{multicol}

\usepackage{textcomp}
\usepackage{graphicx}
\usepackage{caption}

\usepackage{soul}\soulregister\cite7\soulregister\ref7\soulregister\pageref7

\usepackage{zed-cm}
\usepackage{csp-cm}

\usepackage{verbatim} 

\usepackage{url}

\usepackage{xcolor}

\newcommand{\purpose}[1]{}
\newcommand{\research}[1]{}
\newcommand{\planning}[1]{}
\newcommand{\note}[1]{}
\newcommand{\margin}[2]{}{}

\usepackage{tabularx}

\usepackage{rotating}

\usepackage{pbox}

\usepackage{footnote}
\makesavenoteenv{tabularx}
\makesavenoteenv{tabular}
\makesavenoteenv{table}

\usepackage{paralist}

\begin{document}
\maketitle

\begin{abstract}
	Ensuring a car's internal systems are free from security vulnerabilities is of utmost importance, especially due to the relationship between security and other
	properties, such as safety and reliability. We provide the starting point for a model-based framework designed to support the security testing of modern cars. We use Communicating Sequential Processes (CSP) to create architectural models of the vehicle bus systems, as well as an initial set of attacks against these systems.
	While this contribution represents initial steps, we are mindful of the ultimate objective of generating test code to exercise the security of vehicle bus systems.
We present the way forward from the models created and consider their potential integration with commercial engineering tools.

\end{abstract}

\section{Introduction}\label{sec:intro}

The security of modern cars has attracted a great deal of media attention in recent years (e.g \cite{thielman_someone_2016}).  As the transportation paradigm shifts towards \emph{autonomy}, many concerns have been raised about the in-car computer systems' ability to safeguard the behaviour of cars --- and, ultimately, the safety of drivers, passengers and pedestrians.  


Several methods are available to test the security of systems against attacks, one of which (and possibly the most well-known) is \emph{penetration testing}. In it, 
highly-skilled security professionals mimic the actions of an attacker, whose objective is the achievement of unauthorised goals (e.g. data exfiltration or command spoofing) on the target system.  
Although penetration testing is widely deployed by companies seeking to obtain an understanding of the real security of their systems and networks, it is not a perfect solution.
For example, it carries a high cost, it requires a high degree of ability and knowledge, and it is primarily applicable to already deployed systems --- it is a \emph{reactive}, rather than a \emph{preventive} security testing approach.

While we recognise penetration testing's role, we believe that there is the potential for other security testing approaches to play a complementary role, with one candidate being
the testing of a system's security from models --- in short, \emph{model-based security testing}. 
Thus, the focus of this paper, which builds upon the initial work described in~\cite{dossantos_formal_2016}, is on answering \emph{how might a model-based testing (MBT) framework support the automatic security testing of a modern car's subsystems?} This work falls into the category of formal and model-driven approaches to engineering safety-critical and security-critical systems.


Our motivations for answering this question are manifold.  First, it is consistent with the automotive industry's approach to product development, which is characterised by an emphasis on models, with the actual implementation often delegated to third parties.  Second, the distributive nature and complexity of the systems involved require a testing approach reusable across all the different systems.  Finally, initiatives like OpenCar2X~\cite{laux_demo_2016} indicate a momentum towards the creation of in-car apps by various developers (in a similar way to the smart phone eco-system) --- these apps will inevitably have to be tested for security. 



The remainder of this paper is organised as follows. Section~\ref{sec:motivation} details the motivations behind our research line and provides necessary background information on related work. Section \ref{sec:csp} introduces CSP, the formalism backing the work. In the sequence, Section~\ref{sec:thr-mdl} defines the threat model from which security attacks to the system will be derived.  Next, Section~\ref{sec:arc-mdl} presents the core of the work: the definition of the in-vehicle architecture and attacks in a formal fashion.
Finally, Section~\ref{sec:conc} concludes the paper, discussing our results in context and paving the way ahead in the research project.
\section{Motivation and Background}\label{sec:motivation}


Cars are very complex entities. This complexity begins with architecture design, which is formed by a large number of inter-connected embedded systems --- also known as \emph{electronic control units} (ECUs).  An average modern passenger vehicle typically has 40 or more ECUs~\cite{miller_survey_2014}.  Depending on their application, ECUs can have strict networking requirements.  As a result, various kinds of networks are available.  For instance, a low-latency network (e.g. the Controller Area Network) is more suitable for real-time applications, whereas a high-bandwidth network (e.g. MOST) is more suitable for in-car entertainment applications. 



\begin{figure}[t]
 \centering
 \includegraphics[scale=0.5]{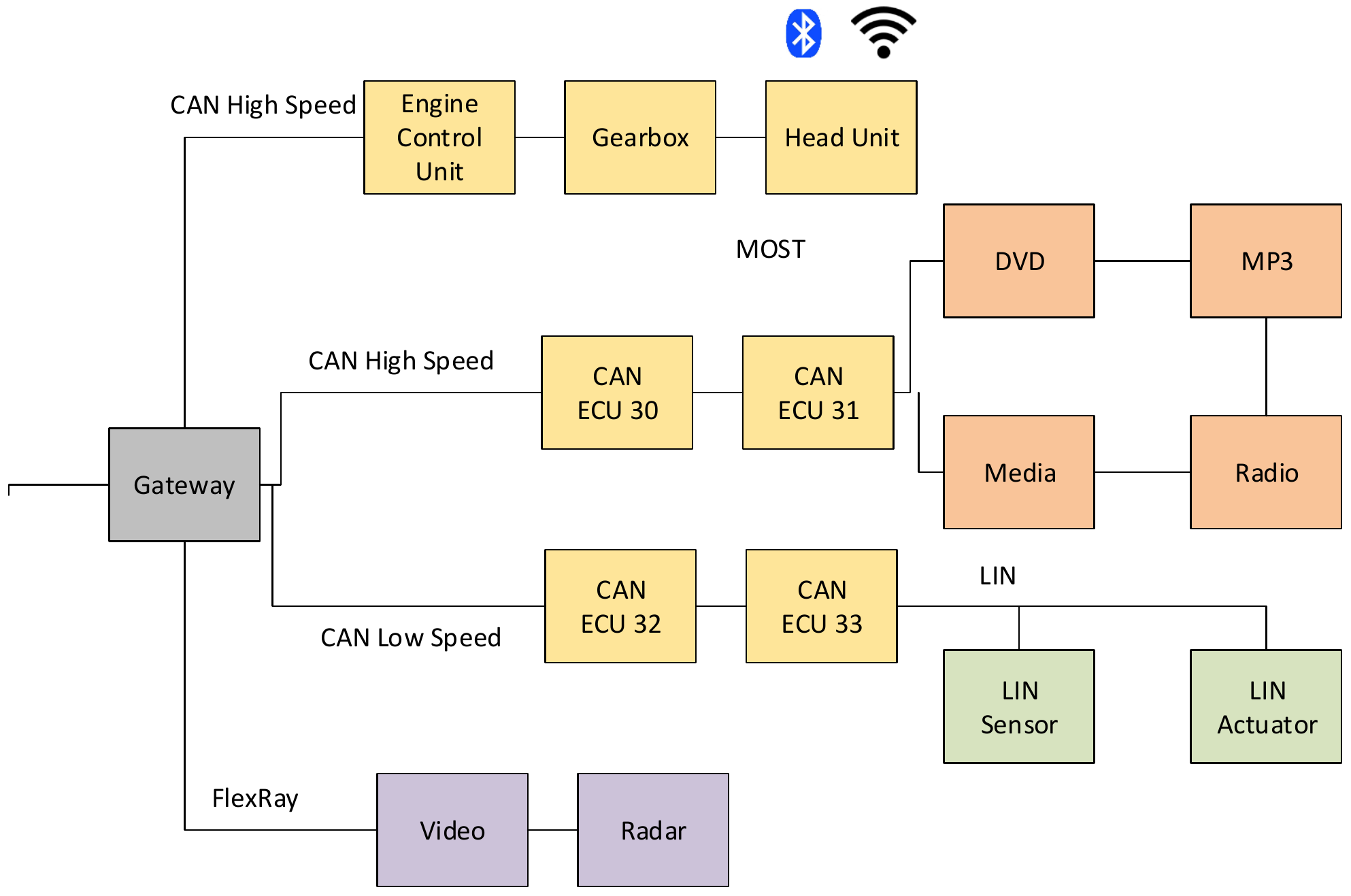}
 \caption{Automotive bus systems in a generic architecture}
 \label{fig:in-vehicle-arch}
 \end{figure}

Figure~\ref{fig:in-vehicle-arch} illustrates a simplified 
version of an automotive architecture, which is based on the topology of bus systems \cite{sommer_vehicular_2015}. Realisations of this architecture can vary, with respect to the number of components, as well as, the form of connection between components.
Various bus systems are connected to different networks. Gateways manage communication between different networks. Each network type provides its own set of requirements (e.g. latency, transmission speed, etc.), as well as has its own frame structure. The Controller Area Network (CAN) is responsible for safety and non-safety functions of the vehicle. 
It should be noted that FlexRay intends to replace CAN as standard for safety communication~\cite{waraus_steerbywire_2009}.

Sensors and actuators can either be connected to a bus system directly or attached to their own Local Interconnect Network (LIN). 
Figure~\ref{fig:in-vehicle-arch} shows a centralised architecture with one gateway.  Decentralised architectures with two gateways
are also possible. 

%
%
%

With regards to the security of cars, early academic studies have shown that cars can be exploited by both physical~\cite{DBLP:journals/ress/HoppeKD11,koscher_experimental_2010} and wireless~\cite{checkoway_comprehensive_2011,DBLP:conf/woot/FosterPKS15} channels.  Physical attacks typically take the form of packet injections directly on the internal network (e.g. the aforementioned CAN) 
via the OBD-II port, whereas wireless attacks typically exploit vulnerabilities in the implementation of mostly proprietary authentication protocols.  Researchers have been able to achieve a range of unauthorised goals, including installation of covert surveillance capabilities within the vehicle, remote data exfiltration and remote control. 
Although all vehicles share the same basic principles and communication standards, it seems apparent that implementation and architectural differences play a vital role to determine a car's security level~\cite{miller_survey_2014}. 

To date, most 
studies have focussed on finding network issues. Less attention, however, has been given to security issues that are brought about by data~\cite{petit_revisiting_2014}. Vehicles nowadays consume a vast amount of data from a variety of sources (e.g.~data collected by internal and external sensors, and from online databases).  Further, within the vehicle, data needs to be captured, processed and stored~\cite{petit_revisiting_2014}.  All of these steps must happen securely to provide the vehicle with higher assurance levels.  
We consider this data lifecycle in this paper.


With regards to formal methods, there has been little application of them in the automotive domain broadly, and even less for security specifically. 
Mundhenk \emph{et al.}~\cite{mundhenk_security_2015} use continuous time Markov models to facilitate decision making between different automotive architectures with regard to required security properties. Siegl \emph{et al.}~\cite{siegl_formal_2011} formalise automotive systems requirements by test models. However, their approach is concerned primarily with safety, rather than security requirements. 

Modern cars inherit many of the more generic verification challenges associated with cyber-physical systems (CPSs). Zheng and Julien~\cite{zheng_verification_2015} outline some of the challenges facing verification of cyber-physical systems, noting that CPSs' real-time requirements and traditional large system size make them difficult to be verified by state-of-the-art formal verification tools.

Seshia \emph{et al.}~\cite{seshia_formal_2015} discuss the need for formal methods for semi-autonomous driving. By requiring human driver intervention at certain moments, semi-auto\-nomous cars introduce many conditional specification and verification problems. For instance, it is not easy to determine which situations require human intervention or whether the driver will be able to resume control of the vehicle on time. 


It should be noted that, in the following, we make no distinction between \emph{car} and \emph{vehicle} --- their use is synonymous.  For simplicity, whenever we refer to \emph{automotive system}, we also refer to the connecting networks associated with the given system.

\section{Communication Sequential Processes (CSP)}\label{sec:csp}

Our formal models are presented in the language of Communication Sequential Processes (CSP), a process-algebraic formalism that supports the modelling and analysis of concurrent systems.
In addition we utilise the refinement checker, FDR (Failures Divergences Refinement)~\cite{oxforduniversity_fdr4_2017}.

The choice of CSP can be justified by the systems of systems (SoS) approach to building modern vehicles, which can be modelled in CSP via compositions of processes. In this context, each individual sub-system is modelled as an individual CSP process. Processes, in turn, can exchange messages between each other in a similar way to automotive networks. 
There are precedents for the use of CSP in this respect: it has also been in the verification of properties of car protocols (see, for example, \cite{ran_modeling_2014}).

In the following, we provide a necessarily brief introduction to the language of CSP.

\subsection{Syntax}\label{sec:syntax}

The \emph{alphabet} of a CSP process is the set of events that it is willing to communicate.  An  \emph{event} requires the agreement of both the environment and the communicating process, occurs instantaneously, and is atomic.  The set of all possible events within the context of a particular specification is denoted $\Sigma$.   

The simplest possible process is the one that offers to participate in no events: $STOP$.  Another simple process, $SKIP$, represents successful termination (via the internal event, $\tick$).  The \emph{prefixing} operator, $\then$, allows us to prefix an event $e \in \Sigma$ to any process: the process $e \then STOP$ will refuse to communicate anything other than $e$, after which it will offer no further events.  Recursion allows us to capture infinite behaviour using a finite description. For example, $B = b \then B$ is the process that will communicate $b$ indefinitely. 

\emph{External} (or \emph{deterministic}) choice is denoted by $\extchoice$:  we write $P \extchoice Q$ to describe the process that offers the choice between the initial events of both processes, before behaving as one of $P$ or $Q$.  \emph{Internal} (or \emph{nondeterministic}) choice is denoted by $\intchoice$: the process $P \intchoice Q$ may behave as $P$ or as $Q$, but the environment has no choice over which.  Both forms of choice have an associated indexed form: $\Extchoice i: I @ A(i)$ and $\Intchoice i: I @ A(i)$.

A parallel composition of processes tells us not only which processes are to be combined, but also describes the events that are to be synchronised on.  The \emph{synchronous parallel} operator, $\parallel$, requires component processes to agree on all events.  As an example, the process $A \parallel B$, where $A = a \then b \then STOP$ and $B = b \then B$, will deadlock as the combination cannot agree on the first event.  \emph{Generalised parallel composition}, of the form $A \parallel [X] B$, requires cooperation on the events of $X$; all other events can proceed independently.  For example, in $A \parallel [\{b\}] B$,  the event $a$ occurs without $B$'s cooperation. Some processes do not synchronise on any events; \emph{interleaved} processes are written $A \interleave B$.

\emph{Channels} are the means by which \emph{values} are communicated: $c.x$ is the communication of the value $x$ on the channel $c$.  A natural extension to this is to consider input and output values --- of the form $c?x$ and $c!x$ respectively.  Input along a channel can also be expressed using external choice: we might write $\Extchoice x : X @ c.x \then P~(x)$.

\subsection{Semantics}\label{sec:semantics}
It is often useful to consider a trace of the events which a process can communicate: the set of all such possible finite paths of events which a process $P$ can take is written $\Traces[P]$.  However, it is well understood that traces on their own are not enough to fully describe the behaviour of a process.  For a broader description of process behaviour, we might consider what a process can refuse to do: the refusals set of a process --- the set of events which it can initially choose not to communicate --- is given by $refusals[P]$.  By comparing the refusals set with the traces of a process, we can see which events the process \emph{may} perform: $refusals[A \extchoice B] = \{ \}$ and $refusals[A \intchoice B] = \{\{ \}, \{ a \}, \{ b \}\}$ (assuming $A$ and $B$ as defined above, and $\Sigma = \{a,b\}$).  It follows that  the \emph{failures} of a process $P$ --- written $\Failures[P]$ ---  are the pairs of the form $(t,X)$ such that, for all $t \in \Traces[P]$, $X = refusals[P \mathrm{/} t]$, where $P\mathrm{/}t$ represents the process $P$ after the trace $t$. 

The aforementioned FDR tool compares processes in terms of refinement.  We write $P \refinedby[M] Q$ when $Q$ refines $P$ under the model $M$: $Q$ is `at least as good as' $P$.  If we were only to consider traces, then
\[
P \refinedby[T] Q \iff \Traces[Q] \subseteq \Traces[P]
\]
Similarly, we define failures refinement as
\[
P \refinedby[F] Q \iff \Traces[Q] \subseteq \Traces[P] \land \Failures[Q] \subseteq \Failures[P]
\]

\section{Threat Model}\label{sec:thr-mdl}

We now briefly review the supporting threat model of our work.
For a broader review of the topic, we refer the reader to~\cite{brooks_chapter_2012} and~\cite{mokhtar_survey_2015}.

\begin{enumerate} 
	\item \emph{Vehicle Owner/Driver.}  Car drivers have unlimited access to the vehicle, which, in most cases, is their own property.  They do not usually have knowledge about the internal functioning of a car.
	Some car models allow external apps to interact with the car or to add non-safety related software component to the head unit. 
	Drivers are not limited to the vehicle's legal owner or their relatives, but rather should include everybody allowed to drive the vehicle (e.g. valets, car-sharers, etc.).
	
	\item \emph{Evil Mechanic.}  An evil mechanic has unlimited access to many cars at the same time.  They also have a deep knowledge of cars' internal functioning as well as specialist diagnostics equipment.  As a result, they can easily modify cars' hardware and software components.  In addition, they can also manipulate data stored within the vehicle~\cite{petit_potential_2015}.  Commonly, drivers have no way to tell what modifications their car underwent after being collected from the garage.
	
	\item \emph{Thief}. 
	Thieves might be keen to steal the owner's possessions inside a car, car components, or the car as a whole. 
	A thief can attempt to get access to the bus system, attach a malicious ECU and send fabricated messages possibly opening the doors, evading the immobilizer and starting the engine~\cite{miller_adventures_2013}.  Since many cars share the same technology and car theft can have a high profit margin, acquiring knowledge and tools is worthwhile for professional thieves.  The knowledge and tools might be acquired via reverse engineering or by colluding with evil mechanics.
\planning{look for price of tools and car engineering courses	}
\planning{look for statistics on thieves collusion with mechanics}	
	
	
	\item \emph{Remote Attacker.}
	Some individuals might want to bring havoc to many cars indiscriminately out of various possible motivations, e.g. the seeking of a feeling of power.  To have maximum impact, an attacker could try to use any of the openly accessible interfaces of a car such as cellular connections usually used to connect the car via the internet to a back-end system, wireless connections for Vehicle-to-X communication (IEEE 802.11p), or Bluetooth connecting mobile devices to the head unit (cf. \cite{checkoway_comprehensive_2011,koscher_experimental_2010,miller_survey_2014}).  
\end{enumerate}
%


Our threat model assumes that an attacker has complete access to the wired network of a car in addition to access to public interfaces.  An attacker can modify the network, i.e. remove or add any network connection or ECU.  With respect to data manipulation, if the attacker manages to modify data (software or configuration data) --- as done in the Jeep Cherokee hack~\cite{miller_remote_2015} --- the attacks can cause a change to the built-in functionality


\section{Formal Models of the Automotive Architecture}\label{sec:arc-mdl}

To enable the generation of test cases from models, there is a need to specify the underlying system to be tested in terms of models. 

Our formal architecture consists of modelling individual networks and associated bus systems as CSP processes.  We use the generic architecture of~Figure \ref{fig:in-vehicle-arch} as a basis.  Instead of modelling the generic architecture as a single entity, we decompose it into smaller parts, each of which represent a single network type or system. We are aware that a bus system may be formed by multiple ECUs~\cite{sommer_vehicular_2015}.  However, for the sake of simplicity, we model each bus system as associated with a single ECU only. We use CSP's parallel composition operator ($\parallel$) to denote communication between bus systems and the corresponding network.

\subsection{The system model}

Our system model follows the diagram in Figure \ref{fig:in-vehicle-arch}. We model gateways, networks, and bus systems connected to each network. 

The first process, $GATEWAY$, manages the communication between the networks directly associated to it: two CAN High-Speed networks ($CANHS1$ and $CANHS2$), one CAN Low-Speed network ($CANLS$), and one FlexRay network ($FLEXRAY$). 

Each network has its own CSP process, where associated bus systems are modelled.  Events of the form $gateway\_*$ serve to communicate between the process $GATEWAY$ and individual network processes. External choices ($\extchoice$) split different possible paths (i.e. between gateways or bus systems).  After the $gateway\_*$ event in the network process (e.g. $CANHS1$), there is an external choice between available bus systems within that process.  This multi-level choice is repeated many times in our models. 

\[
GATEWAY = 
\begin{array}[t]{l}
gateway\_canhs1 \then CANHS1 \\
\extchoice \\
gateway\_canhs2 \then CANHS2 \\
\extchoice \\
gateway\_canls \then CANLS \\
\extchoice \\
gateway\_flexray \then FLEXRAY
\end{array}
\]

\[
CANHS1 = 
\begin{array}[t]{l}
gateway\_canhs1 \then \\
\t1 ( \begin{array}[t]{l} 
engine\_cu \then STOP \\
\extchoice \\
gearbox \then STOP \\
 \extchoice \\
 head\_unit \then STOP~)\\
 \end{array}   
 \end{array}
\]
    
\[
CANHS2 = 
\begin{array}[t]{l}
gateway\_canhs2 \then \\
\t1 (~can\_ecu30 \then STOP \extchoice canhs\_most \then MOST~)
\end{array}
\]

\[
CANLS = 
\begin{array}[t]{l}
gateway\_canls \then \\
\t1 (~can\_ecu32 \then STOP \extchoice canls\_lin \then LIN~)
\end{array}
\]

\[
FLEXRAY = 
\begin{array}[t]{l} 
gateway\_flexray \then \\
\t1 (~video \then STOP \extchoice radar \then STOP~)
\end{array}
\]
The MOST and LIN networks do not connect with the main gateway. Their connection to individual ECUs in the CAN High-Speed and CAN Low-Speed networks, respectively, is also represented in our CSP models.  The events $canhs\_most$ and $canls\_lin$ indicate those connecting ECUs.

\[
MOST = 
\begin{array}[t]{l}
canhs\_most \then \\
\t1 ( \begin{array}[t]{l} 
dvd \then STOP \\
\extchoice \\
mp3 \then STOP \\
\extchoice \\ 
radio \then STOP \\
\extchoice \\
media \then STOP~)
\end{array}   
\end{array}
\]
    
\[
LIN = 
\begin{array}[t]{l} 
canls\_lin \then \\
\t1 (~lin\_sensor \then STOP \extchoice lin\_actuator \then STOP~)
\end{array}
\]

\subsection{Modelling attacks}

The capabilities of the attacker are described as channels and are defined in process $Attacker$ below. 

The attacker can spoof commands to bus systems, block communication to a bus system, eavesdrop communication between bus systems, and change the functionality of a bus system. These capabilities are expressed by channels $spoofing$, $block$, $eavesdrop$, and $change\_functionality$, respectively.

Spoofing, blocking and eavesdropping are associated with command-and-control or monitoring software, which are installed or operated by remote attackers, thieves or, even, the vehicle's owner. Change of functionality, on the other hand, requires a more advanced knowledge of the vehicle's internal systems as well as available time, thus it is more likely to be conducted by evil mechanics or vehicle owners. The Jeep Cherokee vulnerability discovered by Miller and Valasek is an example of spoofing combined with change of functionality \cite{miller_remote_2015}.

We also define the set $All\_Buses$, which include all bus systems in the vehicle. However, any subset of the bus systems can be used, if required. Variable $c$ of channel $spoofing$ indicate a command to be spoofed to the bus system.
\[
Attacker = \\
\hspace{0.25cm}spoofing?b : All\_Buses?c:\{0..10\} \then Attacker \\
\hspace{0.25cm}\extchoice block?b : All\_Buses \then Attacker\\
\hspace{0.25cm}\extchoice eavesdrop?b : All\_Buses \then Attacker\\
\hspace{0.25cm}\extchoice change\_functionality?b : All\_Buses \then Attacker
\]

Integration between system and attack models is done via  alphabetised parallel composition, as defined by process $Attack1$ below. This type of parallel composition demands the definition of the  events that we want to execute. In the example below, this is done by the set $AttackPath$, which restricts all executable events to $gateway\_canhs1$, $engine\_cu$, and $spoofing$. 

\[
Attack1 = Attacker [ AttackPath || AttackPath ] GATEWAY\\
AttackPath = \{gateway\_canhs1, engine\_cu, spoofing.engine\_cu.2\}
\]

A benefit of such models is that they allow a fine-grained definition of what systems to test and what attacks to test against. Moreover, the capabilities of each threat actor can be represented by different test models. Even though process $Attack1$ uses the whole system model (process $GATEWAY$) as input, only a few events of it are actually considered (via the $AttackPath$ set). Replacing $GATEWAY$ by another subsystem can narrow down the scope of the test models even further. 
\section{Final Considerations and Future Work}\label{sec:conc}
We have created formal models of architectures of the networks and bus systems commonly found in a modern car. We then combine these models with attack models, thus enabling the automatic generation of tests to exercise the behaviour of car systems against attacks. 

Our work also addresses the problem of scalability of security testing in distributed environments. Each of the involved systems must have its security tested. Consequently, the use of automatic tools as well as an underpinning formal model to guide the process is mandatory. Moreover, another advantage of our formal model is that it allows network parameters (e.g. latency) to be modelled alongside the network topology. 
We believe this feature can be useful to test the availability of components --- e.g. sensors, actuators and ECUs --- while the network they are plugged to is under attack. Remote attackers may selectively shut components off or trigger on-demand malicious commands ---, consequently, there is a need to ensure redundancy mechanisms work as expected.

The next step in our work plan is to implement a tool to generate executable test cases from the test models introduced in this paper. This tool will use system and attack models in CSP as input and, with the aid of FDR, will generate traces for each execution of the input process. We anticipate that each trace generated by FDR will correspond to a different test case, which can further vary in terms of process variables. To interface with automotive engineering software, test cases will be generated as XML files or CAPL\footnote{Communication Access Programming Language, a language, similar to C, used in the  programming of automotive bus systems \cite{vector_capl_2017}.} code. 

Furthermore, we also plan to extend the framework to encompass other main sources of threats to the modern car, for instance, attacks from vehicular networks (V2X) and attacks to autonomous vehicles' sensors (e.g. Lidar, cameras, etc). Sensor attacks have been shown in practice (e.g. \cite{shin_illusion_2017}). A formal model of these attacks can also help to systematise the security testing of modern vehicles. Moreover, we will also pursue the integration of our test case generation method into industry-grade automotive engineering tools (e.g. CANoe~\cite{vector_canoe_2017} and PreScan \cite{tassinternational_prescan_2017}). Subsequent work will include the incorporation of our method onto a wider automotive testing methodology and its expansion to other kinds of cyber-physical systems with similar requirements (e.g. avionics).

\vspace{0.3cm}
\noindent \textbf{Acknowledgments.} Eduardo dos Santos thank CAPES Foundation (grant no. 1029/13-4) and Keble College for financial assistance during the development of this research. The authors thank the anonymous reviewers for their constructive comments.

\bibliographystyle{eptcs}

\bibliography{bibliography/eduardo,bibliography/dominik,bibliography/andrew}

\end{document}